\documentclass[10pt,english,journal]{IEEEtran}
\usepackage[T1]{fontenc}
\usepackage[latin9]{inputenc}
\usepackage{geometry}
\usepackage{amsmath}
\usepackage{xpatch}
\usepackage{amssymb}
\usepackage{esint}
\usepackage{mathtools}
\usepackage{amsthm}
\usepackage{nicefrac}
\usepackage{bigints}
\usepackage{mathtools}
\usepackage{blkarray, bigstrut}
\usepackage{physics}
\usepackage{calligra}
\usepackage{graphicx}
\usepackage{epstopdf}
\usepackage{dsfont}
\usepackage{array,ragged2e}
\usepackage{enumitem}
\usepackage{etoolbox}
\usepackage{babel}
\usepackage[nopar]{lipsum}
\usepackage{psfrag}
\usepackage[ruled,lined,linesnumbered]{algorithm2e}
\usepackage{algorithmic}
\usepackage{algorithm2e}
\usepackage{balance}
\usepackage{float}
\usepackage{hyperref}

\usepackage{subcaption}

\SetKw{KwBy}{by}

\makeatletter
\newcommand{\removelatexerror}{\let\@latex@error\@gobble}
\makeatother

\graphicspath{ {Figures/} }

\xpatchcmd{\proof}{\hskip\labelsep}{\hskip5\labelsep}{}{}  
\makeatletter
\xpatchcmd{\proof}{\@addpunct{.}}{\@addpunct{:}}{}{}
\makeatother

\renewcommand\[{\begin{equation}}
\renewcommand\]{\end{equation}} 
\usepackage{listings}
\usepackage{fancyvrb}
\usepackage{framed}

\usepackage{courier}
\usepackage[usenames,dvipsnames,table]{xcolor}

\definecolor{dkgreen}{rgb}{0,0.3,0}
\definecolor{gray}{rgb}{0.5,0.5,0.5}

\DeclarePairedDelimiter\floor{\lfloor}{\rfloor}

\makeatletter
\newcommand*{\rom}[1]{\expandafter\@slowromancap\romannumeral #1@}
\makeatother

\usepackage{siunitx}
\usepackage{tabu}
\usepackage{booktabs}
\usepackage{multirow}
\usepackage{capt-of}
\usepackage{array}
\usepackage{arydshln}
\newcommand{\comment}[1]{}
\newcommand{\change}[1]{{\color{black} {#1}}}

\begin{document}

\title{


Explanation-Guided Deep Reinforcement Learning for Trustworthy 6G RAN Slicing

}

\author{\IEEEauthorblockN{
Farhad Rezazadeh\IEEEauthorrefmark{1}\IEEEauthorrefmark{2},
Hatim~Chergui\IEEEauthorrefmark{3},
and Josep Mangues-Bafalluy\IEEEauthorrefmark{1}\\
}
\IEEEauthorrefmark{1}\normalsize{}Centre Tecnol\'ogic de Telecomunicacions de Catalunya (CTTC), Barcelona, Spain\\
\IEEEauthorrefmark{2}Universitat Polit\'ecnica de Catalunya (UPC), Barcelona, Spain\\
\IEEEauthorrefmark{3}i2CAT Foundation, Barcelona, Spain\\
{\normalsize{}Contact Emails:  \texttt{\{name.surname\}@cttc.es}, \texttt{chergui@ieee.org}
}
}

\maketitle

\begin{abstract}

The complexity of emerging sixth-generation (6G) wireless networks has sparked an upsurge in adopting artificial intelligence (AI) to underpin the challenges in network management and resource allocation under strict service level agreements (SLAs). It inaugurates the era of massive network slicing as a distributive technology where tenancy would be extended to the final consumer through pervading the digitalization of vertical immersive use-cases. Despite the promising performance of deep reinforcement learning (DRL) in network slicing, lack of transparency, interpretability, and opaque model concerns impedes users from trusting the DRL agent decisions or predictions. This problem becomes even more pronounced when there is a need to provision highly reliable and secure services. Leveraging eXplainable AI (XAI) in conjunction with an explanation-guided approach, we propose an eXplainable reinforcement learning (XRL) scheme to surmount the opaqueness of black-box DRL. The core concept behind the proposed method is the intrinsic interpretability of the reward hypothesis aiming to encourage DRL agents to learn the best actions for specific network slice states while coping with conflict-prone and complex relations of state-action pairs. To validate the proposed framework, we target a resource allocation optimization problem where multi-agent XRL strives to allocate optimal available radio resources to meet the SLA requirements of slices. Finally, we present numerical results to showcase the superiority of the adopted XRL approach over the DRL baseline. As far as we know, this is the first work that studies the feasibility of an explanation-guided DRL approach in the context of 6G networks.
 
\end{abstract}


\section{Introduction}

\IEEEPARstart{6}{G} slicing is a disruptive technology that hosts multiple virtual networks with different quality of service (QoS) and tailored to meet the dynamic requirements of vertical services under stringent service level agreements (SLAs). The logically-isolated network instances are envisioned as a vital and integral enabler in future 6G networks, which will underpin a wide range of micro and macro services. Thereupon, it leads management and orchestration (MANO) operations to significant challenges while increasing complexity. Specifically, the quest for automation solutions in 5G/6G has aroused intensive research on the applications of AI and ML. Notably, novel practices are required to adopt dominant AI methods and provide robust models in network slicing \cite{Globe_far} to meet specific service requirements. Additionally, as stated in the European Commission's technical report on "Ethics guidelines for trustworthy AI"\footnote{\url{https://digital-strategy.ec.europa.eu/en/library/ethics-guidelines-trustworthy-ai}}, AI solutions should be trustworthy. In this intent, a significant challenge in fulfilling  AI-driven  B5G/6G is to guarantee the reliability and transparency of complex AI models, which are inherently black boxes with limited insight into the processes that occur from input to output and the reasoning behind predictions and decisions. Nevertheless, the promising XAI features, such as interpretability and openness can ease this concern and enhance trust among stakeholders in the network slicing ecosystem.

Several organization initiatives and academic research works strive to adopt AI techniques in telecommunications to achieve fully-automated service provisioning in 5G and beyond. Within the network slicing literature, there is a limited work pertains to leverage the application of XAI and XRL. The 3GPP RAN3\footnote{\url{https://www.3gpp.org/3gpp-groups/radio-access-networks-ran/ran-wg3}} has studied a functional framework for AI/ML functions in 5G architecture, and the ITU-T SG13 ML5G\footnote{\url{https://www.itu.int/en/ITU-T/focusgroups/ml5g/Pages/default.aspx}} has developed a technical specifications framework for ML in future networks. The Next-Generation Mobile Networks (NGMN)\footnote{\url{https://www.ngmn.org/}} Alliance has published an End-to-End (E2E) architecture framework for 6G use cases, and the O-RAN\footnote{\url{https://www.o-ran.org/}} Alliance is focusing on evolving 3GPP access with openness, virtualization, and AI-enabled RAN. The European Telecommunications Standards Institute Experiential Networked Intelligence (ETSI ENI)\footnote{\url{https://www.etsi.org/technologies/experiential-networked-intelligence}} is developing a cognitive network management architecture that utilizes AI techniques and aims to provide fully-automated service provision, operation, and assurance for all networks, including 5G. ETSI ENI is also working on proof of concepts to demonstrate the potential of AI solutions for network operations, such as optimized slice management and resource orchestration. The ETSI Zero-Touch Service Management (ETSI ZSM)\footnote{\url{https://www.etsi.org/technologies/zero-touch-network-service-management}} is investigating automation challenges faced by operators and vertical industries in the deployment of 5G and network slicing, and working on a new architecture framework for closed-loop automation and AI/ML algorithms.

The authors in \cite{XAI-Barnard} has invoked an XAI methodology to demystify AI model behaviour for short-term resource reservation (STRR) problem within network slicing. The experiments reveal the important trends about the real-time decisions of the model. Saad \emph{et al.}  \cite{XAI-Saad} has proposed a  framework that combines XAI and federated learning (FL) to predict and interpret the slices' latency key performance indicator (KPI). The framework constitutes building a DNN model to predict the KPI in a federated manner and then incorporating various XAI models such SHAP to provide more transparency and explainability to the predictions. The experimental results show the efficiency of the proposed deep learning (DL)-based scheme. In \cite{XAI_Rahman}, the authors proposed a framework based on network slicing and a software-defined network (SDN) at the edge leveraging explainability and semantics into existing DL models to allow human domain experts on COVID-19 to gain insight and semantic visualization of key decision-making processes. The considered proof-of-concept (PoC) shows promising results. The author in \cite{XAI_Guo} has outlined the fundamental techniques of XAI in 6G wireless networks, including the primary public and legal motivations, various definitions and concepts related to explainability, and different algorithms used in XAI.

From this state-of-the-art (SoA) overview, incorporating these initiatives can adopt the AI and XAI mechanisms into the network slicing to address the absence of an effective and interpretable solution in 6G RAN. In this regard, XRL is a reliable and trustworthy machine learning (ML) technique where the reinforcement learning (RL) agent interacts in real-time with slice instances to generate the dataset on the fly and continuously improve the learning performance. It is assisted with DL in DRL algorithm to overcome the curse of dimensionality and challenges of large state spaces. However, to fully harness the potential of the DRL algorithm, it is necessary to streamline the complex relationship between states and actions to avoid conflicts in action selection, which can hinder the successful implementation of automated network slicing. The XAI can help DRL to identify more relevant state-action pairs and explain which states or inputs have the greatest positive impact on actions or decisions.

To validate our proposed XRL approach and form these principles into one cohesive design, we target a resource allocation optimization problem where the explanation-guided DRL agent learns to assign the optimal available radio resources to the slices aiming to minimize the transmission latency and fulfill the SLA requirements. In this paper, we present the following contributions:
\begin{itemize}
    \item The RAN resource allocation problem is framed as an optimization problem aiming to minimizing the SLA violation.
    \item We introduce a novel intrinsic interpretable multi-agent DRL approach to allocate optimal radio resources to slices while enhancing decision-making transparency.
    \item We design an XRL pipeline assisted by SHAP importance values and an entropy mapper mechanism to guide the DRL agent in reducing the uncertainty of selected actions across various network states.
    \item We present AI, XAI, and network analysis to showcase the superiority of the proposed explanation-guided DRL approach compared to the RL baseline.
\end{itemize}

\begin{table}[!t]
\caption{\change{Notation Table}}
\label{tab:model_var_and_par}
\centering
\begin{tabular}{@{}lc@{}}\toprule

\textbf{Notation} & \textbf{Description} \\ \midrule
$\mathcal{B}$ & gNB\\ \hdashline
$\mathcal{I}$ & Set of slices\\ \hdashline
$C_b$ & Capacity of gNB $b\in\mathcal{B}$\\ \hdashline
$\Lambda_i$ & Latency requirement\\ \hdashline
$\lambda_i$ & Throughput requirement\\ \hdashline
$\mathcal{T}$ & Set of decision intervals\\ \hdashline
$\epsilon$ & Duration of decision intervals $t\in\mathcal{T}$\\ \hdashline
$a_{i,b}^{(t)}$ & PRB allocation\\
\hdashline
$\sigma_{i,b}^{(t)}$ & Average SNR experienced by the users\\ \hdashline
$\varphi_{i,b}^{(t)}$ & Instantaneous traffic demand of the user\\ \hdashline
$d_{i,b}^{(t)}$ & Dropped traffic\\ \hdashline
$\iota$ & Minimum PRB allocation\\ 
\hdashline

\change{$\gamma$} & \change{Discount factor}\\ 
\hdashline
\change{$\xi$} & \change{Learning rate}\\ 
\hdashline

\change{$\beta_{i}$} & \change{Experience buffer}\\ 
\hdashline

\change{$\theta_{i}^{(t)}$} & \change{Online network parameter}\\ 
\hdashline

\change{$\tilde{\theta}_{i}^{(t)}$} & \change{Target network parameter}\\

\bottomrule
\end{tabular}
\end{table}

\section{Explainability in RL}
\label{XRL-description}
\begin{figure}
\centering
\includegraphics[scale=0.125]{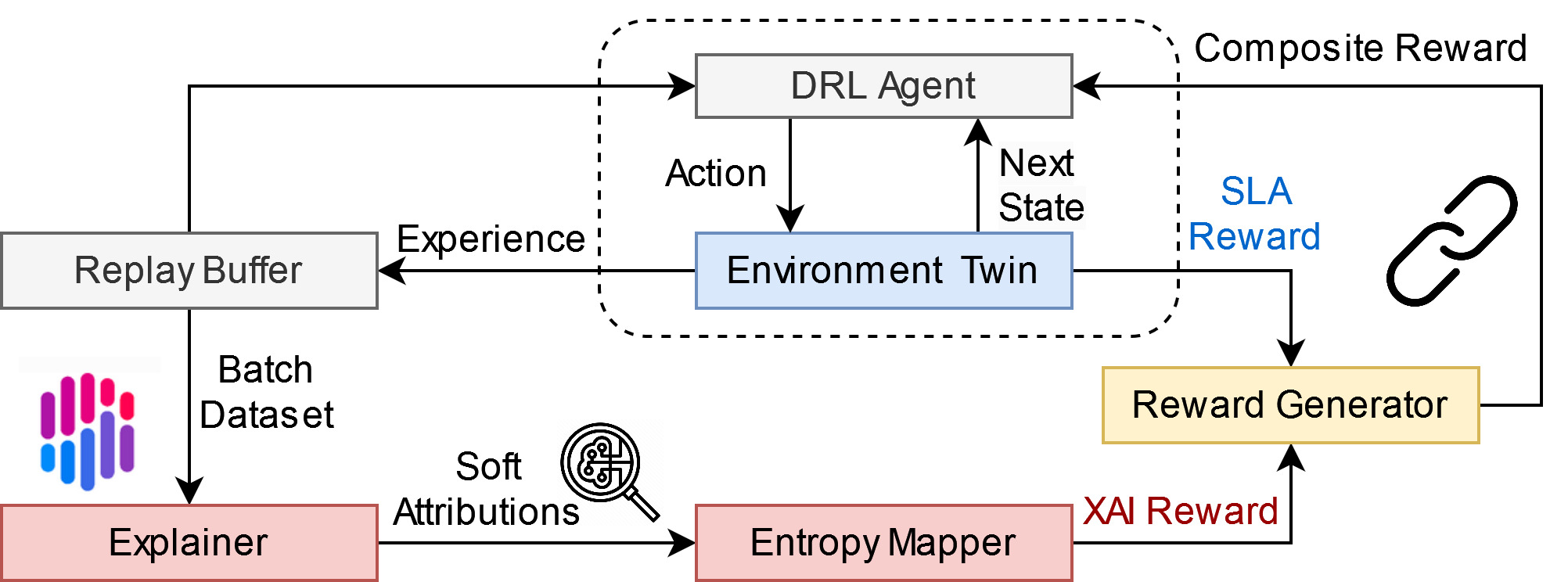}
\caption{\small Explainable DRL workflow.}
\label{fig:XDRL_architecture}
\end{figure}

Despite the promising results and performance, there is a concern about the essence of the deep neural network (DNN)-based DRL, which are deemed as opaque models. This issue is crucial given that high reliability and security are required in realistic network and could impede users from trusting the trained agents and predicted results in the network slicing ecosystem. Motivated by explanation-guided learning (EGL), we consider an intrinsic interpretability approach in the training phase of the proposed XRL agent whose training workflow is illustrated in Fig~\ref{fig:XDRL_architecture}. 

The accuracy and reliability of estimating state-action pairs in DRL are hampered by sparse rewards and a lack of interpretability. As formulated in Section \ref{sec: egl}, this challenge can be addressed when the DRL agent saves its experiences and observations temporarily in a replay memory/buffer after interacting with the Environment Twin (a simulation of the real environment), which is continuously updated. Then, an Explainer uses the SHAP importance values to generate a probability distribution over a batch of state-action data by applying the softmax function to the SHAP values. An Entropy Mapper then calculates the entropy, which reflects the uncertainty of the selected action given the input state. The inverse of the maximum entropy value is then used as the XAI reward. In Section~\ref{Evaluation-results}, we demonstrate that the composite reward, which is the sum of the SLA reward (based on whether the SLA requirements are met or violated) and the XAI reward, reduces the uncertainty of state-action pairs and encourages the agent to choose the best actions for specific network states. This approach helps to clarify the learning process while guiding the learning towards making explainable decisions for a given state.

\section{Network and Problem Formulation}
\subsection{System Model}

\begin{figure}[t]
\centering
\includegraphics[scale=0.6]{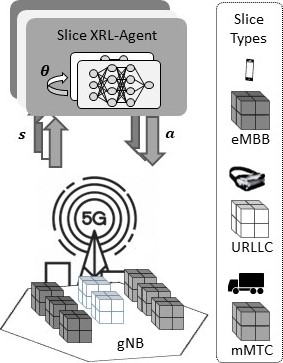}
\caption{\small Generic XRL architecture for RAN slicing.}
\label{fig:xrl-network-slicing}

\end{figure}

In Figure~\ref{fig:xrl-network-slicing}, we consider a radio access network consisting of a base station (BS) $b$ with capacity $C_b$, i.e., a discrete number of physical resource blocks (PRBs) with a fixed bandwidth and also deployed a set of slices $\mathcal{I}$. The available resources are divided into subsets of PRBs and dynamically allocated to each network slice in accordance with their real-time traffic demands and SLA requirements. The variables and parameters are summarized in Table~\ref{tab:model_var_and_par}.

Let us assume a system that operates in time slots, where time is divided into "decision intervals" denoted by $t \in \mathcal{T} = { 1,2,\dots,T}$. In this system, decisions about PRB allocation can only be made at the start of each decision interval. The length of each decision interval, denoted by $\epsilon$, can be determined based on the policies of the infrastructure provider and can range from a few seconds to several minutes.

\subsection{Resource Allocation Formulation}
We see the allocation of radio resources to end-users as a two-step process, as described in~\cite{foukas_orion}. First, once network slices are admitted into the system, the infrastructure provider schedules the assignment of radio resource slots for each tenant. Then, based on the available resources, each tenant may choose to implement their own scheduling solutions for their end-users, based on their specific needs and requirements~\cite{net-Slicing}.

In light of the various user-to-base station association and scheduling algorithms available for allocating resources to end-users~\cite{Spatial_Loads}, our focus is on correctly and fairly dimensioning the allocation of resources between slices, rather than addressing the issue of intra-slice scheduling. To achieve this goal, we utilize the variable $a_{i,b}^{(t)}$ to indicate the decision regarding the allocation of PRBs to the $i$-th slice under the $b$-th BS during the $t$-th decision time interval. Moreover, we employ $\sigma_{i,b}^{(t)}$ to denote the signal-to-noise ratio (SNR) value that represents the wireless channel's quality, which is averaged over a decision time interval $\epsilon$, as well as the end-users of the $i$-th slice attached to the $b$-th BS. Similarly, $\varphi_{i,b}^{(t)}$ is used to indicate the aggregate downlink traffic demand that is generated by the users of the $i$-th slice who are located in the coverage area of the $b$-th BS within the $t$-th time interval.
All together, we formulate the local optimization task as: \\

\noindent \textbf{Problem}~\texttt{RAN Slice Resource Allocation}:
\label{prob:RAN_Slicing}
\begin{flalign}
\text{min} &  \lim\limits_{{T\rightarrow\infty}} \sum\limits_{t = 1}^T \mathbb{E}\left [\sum\limits_{i\in\mathcal{I}} d_{i,b}^{(t)} \right ] \hspace{-3cm}& &\\
\noindent\text{subject to:} \hspace{-1cm}& & &\nonumber\\
& E_{i,b}^{(t)}  \leq \Lambda_i, \hspace{-0.7cm}& \forall t \in \mathcal{T}, \forall i\in\mathcal{I}, \forall b\in\mathcal{B};&\\
& \sum\limits_{i\in\mathcal{I}} a_{i,b}^{(t)} \leq C_b, &\forall t \in \mathcal{T}, \forall b \in \mathcal{B};&\\
& a_{i,b}^{(t)}\in\mathbb{Z}_+, d_{i,b}^{(t)}\in\mathbb{R}_+, &\forall t \in \mathcal{T}, \forall i\in\mathcal{I},\forall b\in\mathcal{B};&
\end{flalign}
where $E_{i,b}^{(t)} = \mathbb{E}\left [ \frac{\varphi_{i,b}^{(t)}}{\Gamma \left(a_{i,b}^{(t)},\sigma_{i,b}^{(t)} \right)+d_{i,b}^{(t)}}\right ]$
represents the expected incurred latency. The function $\Gamma(a, \sigma)$ is used to convert the PRB allocation $a$ into an equivalent transmission capacity with respect to the channel quality $\sigma$. 
The traffic demand during a specific decision interval may be partially satisfied due to an inaccurate estimation of PRB allocation, resulting in additional transmission latency as the traffic waits in the queue at the base station. In this intent, we use variable $d_{i,b}^{(t)}$ as a deficit value that refers to the unserved amount of traffic, i.e., dropped traffic within the agreed slice latency threshold $\Lambda_i$.
\begin{table*}[]
\caption{DRL Parameters}
\label{T2}
\centering
\begin{tabular}{|p{7cm}|p{1.5cm}|p{8cm}|}
\hline 
Parameter & Type & Description\tabularnewline
\hline 
\hline 
$s_i^{(t)} = \{ ( \sigma_i^{(t)}, \lambda_i^{(t)}, \nu_i^{(t)} ) \mid \forall i \in \mathcal{I} \},$ & State Space & At a given time $t$, $\sigma_i^{(t)}$ is the average SNR value experienced by the users in the $i$-th slice over a decision time interval. $\lambda_i^{(t)}$ is the total traffic volume generated by the $i$-th slice during this time interval, and $\nu_i^{(t)}$ represents the remaining available capacity after considering previous allocation decisions made by other agents.\tabularnewline
\hline 
 $\mathcal{A} = \{ \iota \cdot k \mid k = \{0,1, \dots, \floor{\frac{C}{\iota}} \} \}$ & Action Space &  We set $\iota$ as the smallest unit of PRB allocation, also known as the \emph{chunk size}. The PRB allocation decisions made by the $i$-th agent must be in multiples of $\iota$, creating a discrete action space. \tabularnewline
\hline 
\[
    r_i^{(t)} =\begin{cases}
    \alpha^{(t)}_{i} -4 \rho^{(t)}_{\text{lower}}  & \text{if} \qquad \alpha^{(t)}_{i} < \rho^{(t)}_{\text{lower}}, \\
    (1- \frac{\alpha^{(t)}_{i}}{\rho^{(t)}_{\text{up}}}) \frac{\alpha^{(t)}_{i}}{\rho^{(t)}_{\text{up}}}  & \text{if} \qquad \rho^{(t)}_{\text{lower}} \leq \alpha^{(t)}_{i} \leq \rho^{(t)}_{\text{up}},\\
    -(\alpha^{(t)}_{i} - \rho^{(t)}_{\text{up}})                  & \text{if} \qquad \alpha^{(t)}_{i} > \rho^{(t)}_{\text{up}}.
\end{cases}\] & Environment Reward & We assess the quality of the action by introducing two variables, $\rho^{(t)}{\text{up}}$ and $\rho^{(t)}{\text{lower}}$, which define the upper and lower bounds of the allocation gap as $\rho^{(t)}{\text{up}} = 2\cdot\Gamma(\iota^{(t)},\sigma_i^{(t)}) $ and $\rho^{(t)}{\text{lower}} = -\Gamma(\iota^{(t)},\sigma_i^{(t)})$\tabularnewline
\hline 
\end{tabular}
\end{table*}
\subsection{DRL-based Solution}
The abovementioned optimization task can be solved by invoking the DRL framework, wherein the state and action spaces as well as the reward are summarized in Table \ref{T2}.

\subsection{Explainability-Guided Learning}
\label{sec: egl}
A novel approach to measure the confidence of DRL decisions is to observe the distribution of state-features SHAP values in the replay buffer dataset. Specifically, the probability distribution of the states-features is generated as,
\begin{equation}
    \label{softAtt}
         p_{l,k}= \frac{\exp \Bigl\{\left| \alpha_{l,k}\right|\Bigr\}}{\sum _{l'=1}^{L}{\exp \Bigl\{\left|\alpha_{l',k}\right|\Bigr\}}},\,l'=1,\ldots,L, 
\end{equation}
where $\alpha_{l,k}$ stands for the SHAP value corresponding to state $l$ of sample $k$ in the replay buffer dataset.
The decision is viewed as certain when high attributions (in absolute value) are more concentrated in some features minimizing thereby the Shannon entropy,
\begin{equation}
    \mathcal{H}_k = - \sum_{l=1}^{L} p_{l,k} \log(p_{l,k}).
\end{equation}

In this respect, we introduce what we call XRL reward, which is defined as the multiplicative inverse of the entropy, i.e.,

\begin{equation}
    r_{\mathrm{XRL}}^{(t)} = \frac{1}{\max_{k}\mathcal{H}_k}
    \label{eq: xrl-reward}
\end{equation}

Finally, the composite reward fed back to the DRL agent is given by,
\begin{equation}
    r_{i,b,c}^{(t)} = {r}_{i,b}^{(t)} + \mu r_{\mathrm{XRL}}^{(t-1)} \label{eq: comp-reward}
\end{equation}

The single agent training procedure is summarized in Algorithm~\ref{algo:DRL}.


\begin{algorithm}[h]
\caption{Single XRL-Agent Resource Allocation}
 \label{algo:DRL}
\scriptsize
\SetAlgoLined

Initialize primary network $\theta$ and target network $\tilde{\theta}$, and  replay buffer $\beta$,

Import network slicing environment (`XRL--v2'),

Initialize action space $\mathcal{A}$ and state space $\mathcal{S}$

t=0

\While {t < max\_timesteps}{
       \eIf{t < start\_timesteps}{
           Initial buffer filling:
           \scriptsize{$a_{i,b}^{(t)}$ = env.action\_space.sample()}
       }{
             Observe state $s_{i,b}^{(t)}$ and select $a_{i,b}^{(t)} \sim \pi(s_{i,b}^{(t)}, a_{i,b}^{(t)})$
         }
       Execute $a_{i,b}^{(t)}$ and observe $s_{i,b}^{(t+1)}$ and $r_{i,b}^{(t)}+\mu r_{\mathrm{XRL}}^{(t-1)}$:
       
      \quad next\_state, reward, done, 
       
       Store new transition ${({s}_{i,b}^{(t)},{a}_{i,b}^{(t)},{r}_{i,b}^{(t)},{s}_{i,b}^{(t+1)})}$ into $\beta_{i,b}$
       
      \If{t $\geq$ start\_timesteps}{
           Sample batch of transitions $\tilde{\beta}_{i,b}$

           Calculate XRL reward $r_{\mathrm{XRL}}^{(t)}$ according to (\ref{eq: xrl-reward})
           
           Compute target Q value
           
          Perform a gradient descent step on:
           \quad $(y_{i,b}^{(t)} - Q(s_{i,b}^{(t)}, a_{i,b}^{(t)}, \theta_{i,b}^{(t)}))^2$
           
           Update target network parameters:
           \quad${\tilde{\theta}_{i,b}^{(t)}} \longleftarrow \tau {\theta}_{i,b}^{(t)} + (1 - \tau){\tilde{\theta}_{i,b}^{(t)}}$
       }
       \If{done}{
      
       obs, done = env.reset(), False
      }
       t=t+1
  }

 \end{algorithm}
 
\section{Performance Evaluation}


\subsection{Network Scenario and Architecture}
After presenting the key components of the proposed framework, the following crucial step is to analyze how the XRL approach works in realistic settings and assess its performance to validate the superiority of the XRL pipeline compared to the RL baseline. To exemplify the XRL solution, we consider an XRL-driven network slicing setup, as showcased by Fig.~\ref{fig:xrl-network-slicing}, and then measure the model's effectiveness. The lack of dynamic traffic steering hinders efficient and effective resource allocation in network slicing. Over a custom gNB simulator~\cite{Specialization_TVT}, we deploy a slicing setup consisting of three slices, i.e., ultra-reliable low latency communication (URLLC), enhanced mobile broadband (eMBB), and massive machine-type communications (mMTC) with different SLA latencies, $\Lambda_i = [10, 40, 20]$ ms, respectively. The simulator includes virtual transmission queues and implements the PHY/MAC/RLC functionalities.
We formulate the radio resource allocation challenge in gNB as an optimization problem, focusing on minimizing the optimal allocated resources and latency to fulfill the SLA. 
The transmission latency is defined as the average time the traffic experiences within a slice to be served within the gNB transmission buffers due to the inter-slice scheduling process. The slice traffic demand is modeled as a Poisson distribution with a mean value of $\lambda_i$, and its instantaneous values are extracted from a Rayleigh distribution with an average of 25 dB. The radio resource allocation for downlink traffic is divided into subsets of physical resource blocks (PRBs). The multi-agent XRL learns dynamically to allocate the optimal PRBs to each network slice based on real-time traffic and SLA requirements. The gNB used in this scenario has a radio capacity of $C=100$ PRBs of a fixed bandwidth, and all the slices are assumed to run on the gNB simultaneously. A minimum resource allocation chunk size of $\iota = 10$ PRBs is set. In this context, we concentrate on proper and fair dimensioning of inter-slice PRB enforcement instead of addressing the intra-slice scheduling challenge.




\subsection{Experiment Parameters}

We leverage a server running a virtual Ubuntu 20.04 to produce the results. The deep neural networks (DNNs) are implemented using TensorFlow-GPU version 2.5.0 and run on the dedicated server with two Intel(R) Xeon(R) Gold 5218 CPUs @ 2.30GHz and two NVIDIA GeForce RTX 2080 Ti GPUs. The framework is implemented in Python, using the OpenAI Gym library~\cite{masssive-slicing} interfaced with the custom gNB simulator. Each Slice XRL-agent is endowed with a double deep Q-network (DDQN) \cite{GAN_Powered}. These agents interact with other agents and the O-DU through the O-RAN E2 interface to collect slice networking data (such as SNR, served traffic, and consumed resources). The agents then enforce the PRB policy decisions made into the gNB slice scheduler. The DNN architecture consists of two fully connected layers with 24 neurons activated by the ReLU function, and the parameters are updated using the Adam optimizer~\cite{ADAM}. The discount factor $\gamma$ is set to 0.99, and the learning rate $\xi$ is set to 0.001. The replay buffer size for each agent is 20000 samples, from which a batch of 32 samples is fetched for each training interval. To deploy the solution in the cloud-native mode \cite{chapter_far-ICC2023}, we utilized a containerized approach\footnote{\url{https://docs.docker.com/compose}} where the cloud server hosts XRL agents and their corresponding modules and communicate with the network slicing through the FastAPI REST API\footnote{\url{https://realpython.com/fastapi-python-web-apis}}.

\subsection{Numerical Results}
\label{Evaluation-results}

\subsubsection{Long-Term Revenue (Average Reward)}

\begin{figure}[t!]
\centering
\includegraphics[scale=0.55]{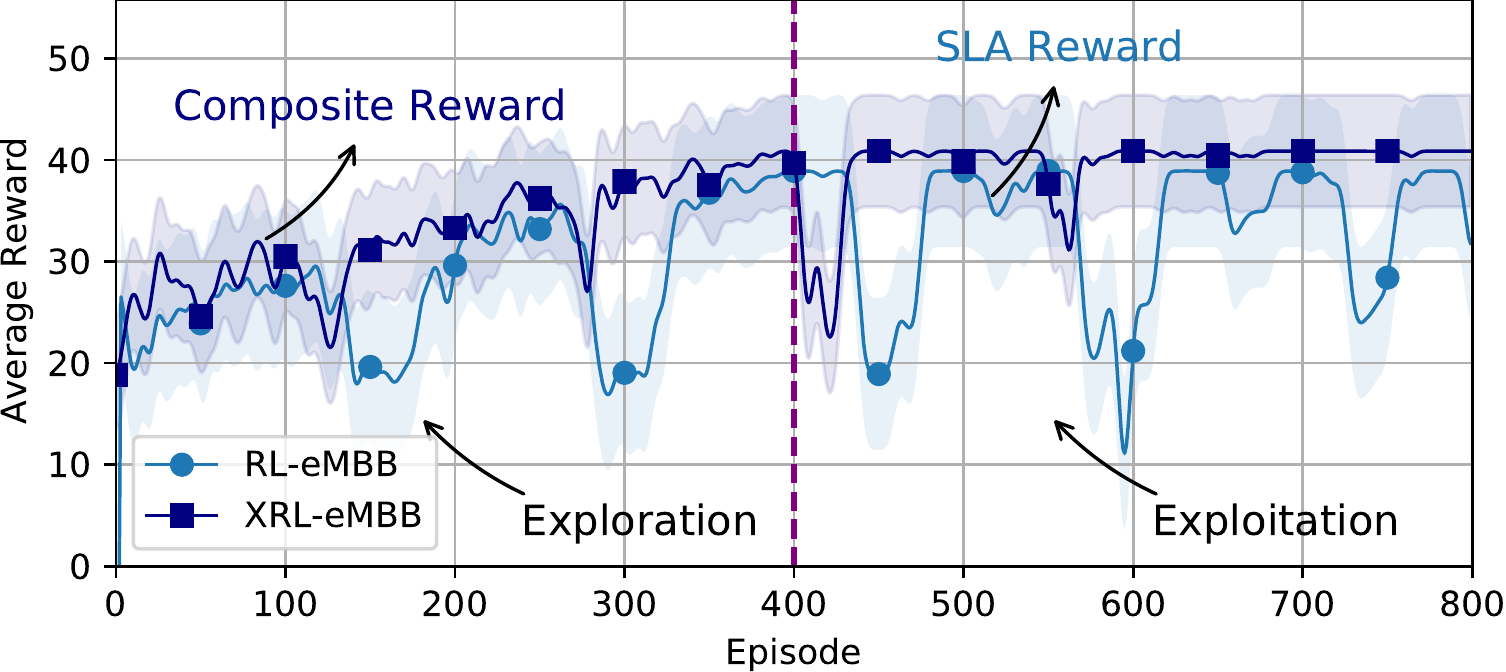}
\caption{\small The convergence curves of the RL and XRL methods.}
\label{fig:rl-xrl-reward}
\vspace{-.5cm}
\end{figure}

As depicted in Fig.~\ref{fig:rl-xrl-reward}, the combination of the SLA reward (RL method) and the proposed XRL reward approach (XAI reward), i.e., composite reward, results in better learning generalization and stability compared to the RL method as the baseline. The explanation-guided action-selection strategy in the XRL approach obtains a higher convergence rate for eMBB slice compared to the RL method. During the initial stages of training, the eMBB agent explores action space (PRBs), which causes high fluctuations in the learning curve, i.e., exploration. However, as training progresses, the agent strives to find an optimal balance between learned decision policies and different network states, i.e., exploitation. It is safe to assert how the proposed SHAP explainer and the entropy mapper work together to analyze the batch dataset, determine the features' relevance, ease the conflict-prone essence of state-action pairs, and guide the agent to make more relevant action decisions for specific states. It is noticed that the curves are smoothed for the sake of visual clarity.
\begin{figure*}
\centering
\includegraphics[scale=0.5]{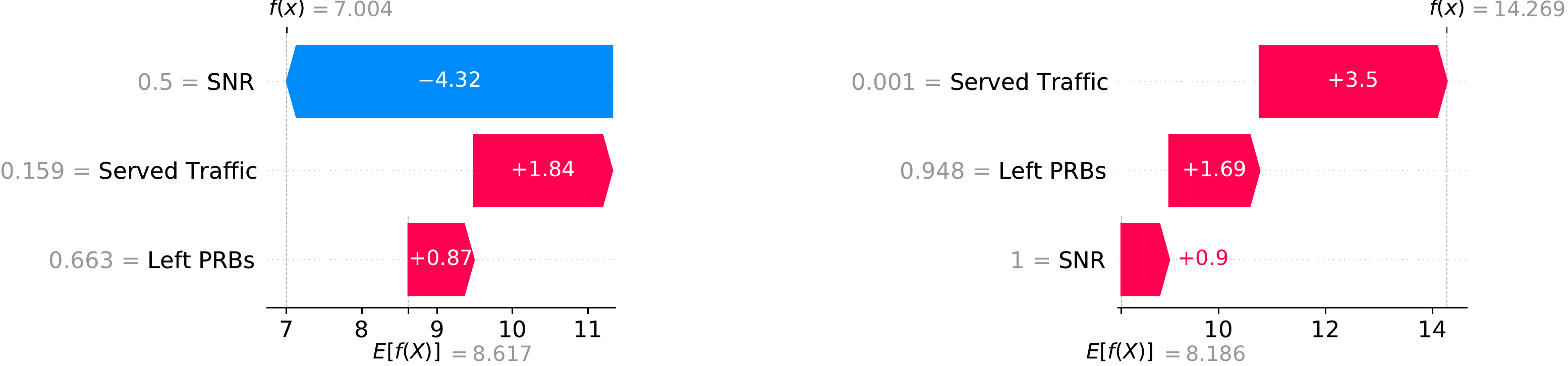}
\caption{\small XAI waterfall plot in the exploration and exploitation phases of URLLC slice.}
\label{fig:SHAP-values}
\vspace{-.3cm}
\end{figure*}
\subsubsection{XAI metric}
In Fig.~\ref{fig:SHAP-values}, $f(x)$ is the predicted action of the XRL agent which is the allocated PRBs to URLLC slice and $\mathbf{E}[f(x)]$ is the expected value, i.e., the mean of all actions. The absolute SHAP value shows us how much a single state affected the action. At the beginning of the training, the agent acts as a Max C/I scheduler, penalizing thereby URLLC users that are experiencing low SNR state (shap value $=-4.32$) and yielding a low PRB provisioning per slice ($7.004$ PRBs). In contrast, at episode $500$, that is during the exploitation phase where the agent has learned the best policy, the agent's action mainly depends on the served traffic (shap value $=3.5$), leading to a higher PRB allocation ($14.269$ PRBs).

\subsubsection{Transmission Latency}

\begin{figure}[t!]
\centering
\includegraphics[scale=0.55]{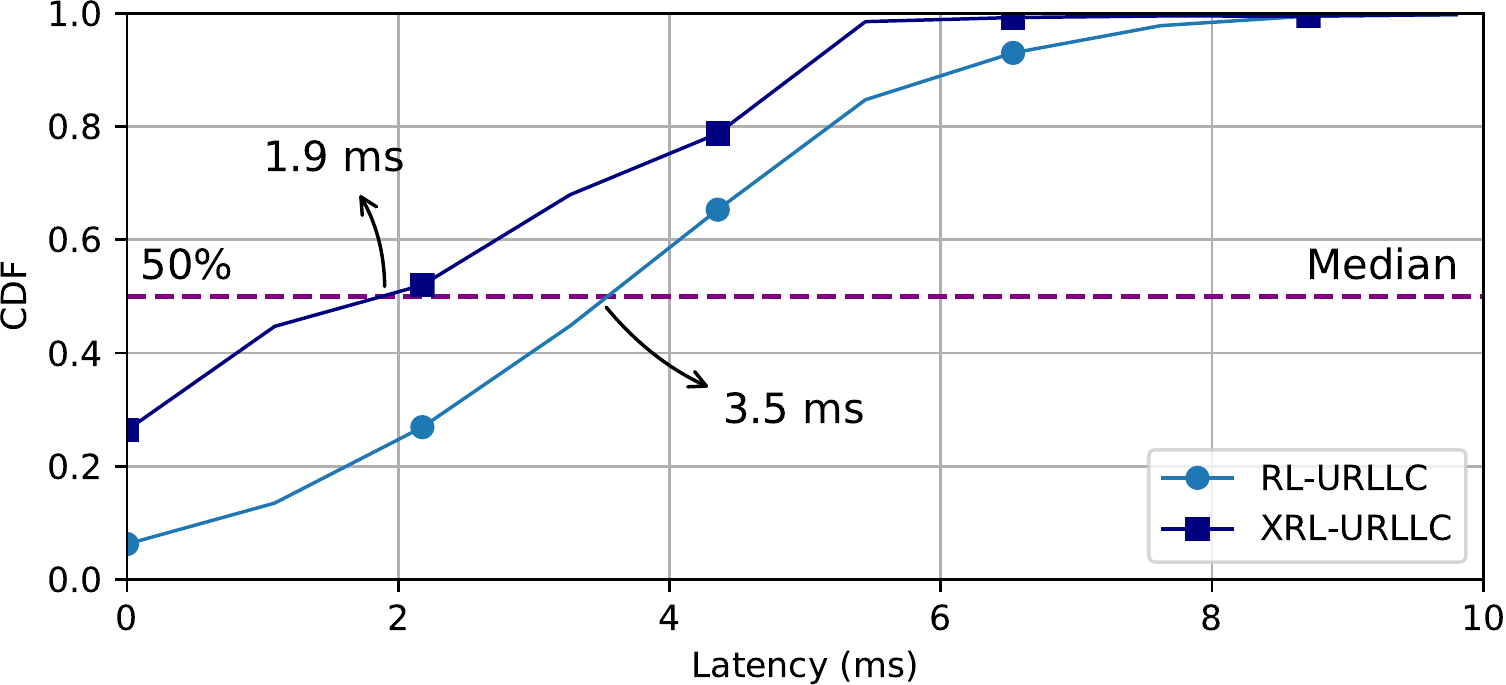}
\caption{\small The CDF of experienced latency within URLLC slice.}
\label{fig:rl-xrl-Latency}
\end{figure}

Fig.~\ref{fig:rl-xrl-Latency} illustrates the cumulative distribution function (CDF) of the latency that URLLC traffic experiences in the gNB transmission buffer, resulting from RL and XRL radio resource allocation policy. The curves measure the probability of incurred URLLC traffic latency within a certain latency deadline during the agent training. The shape of the curves is based on the variability of latency values, which is latency distribution. The results indicate that XRL has a superiority compared to RL solution where \emph{50\%} of perceived latencies by XRL is less than \emph{1.9~ms}, whereas this value for RL agent is \emph{3.5~ms}. It turns out that the proposed XRL approach leads to optimal and adequate PRBs allocation concerning dynamic traffic and resource contention among slices, and then URLLC experiences lower transmission latency. On the other hand, the RL strategy incorrectly balances resource allocation for different network states (state-action pairs), leading to higher latency.

\subsubsection{Dropped Traffic}
\begin{figure}[t!]
\centering
\includegraphics[scale=0.55]{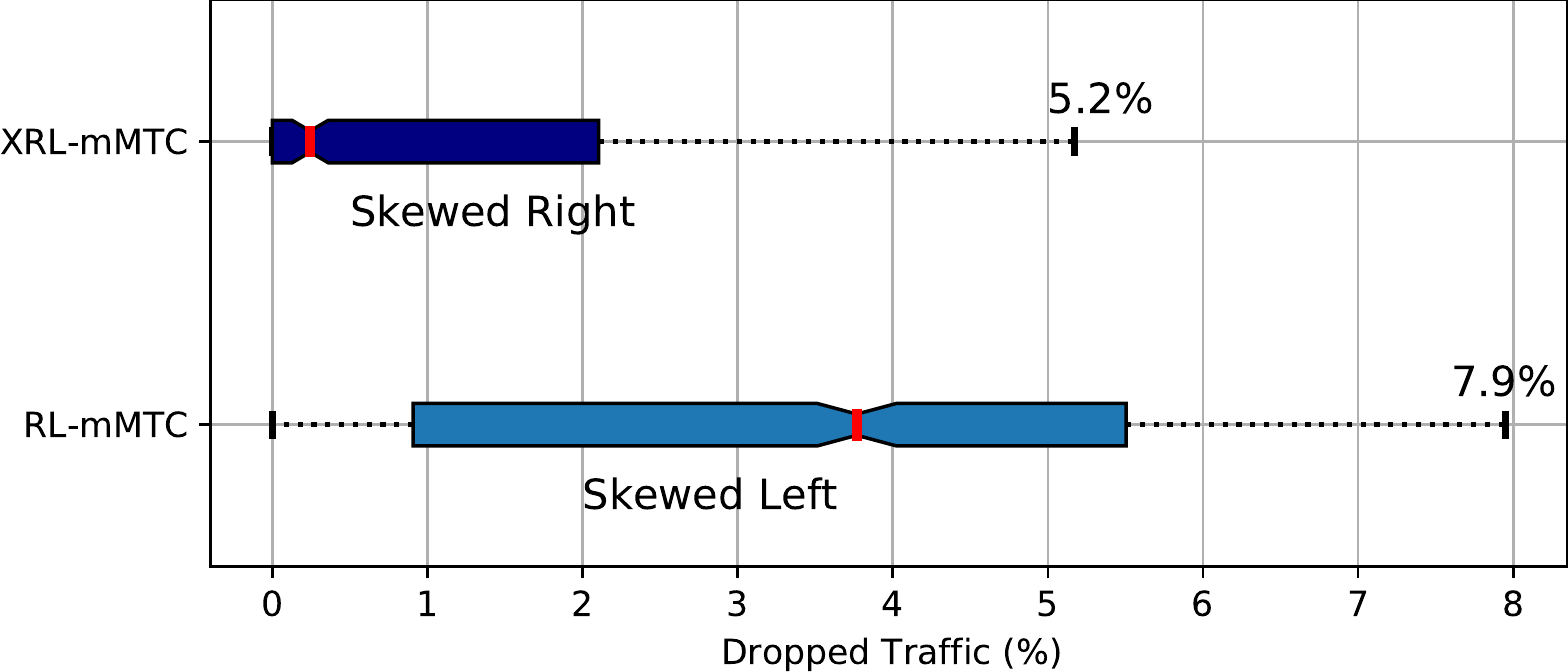}
\caption{\small Performance evaluation for mMTC dropped traffic.}
\label{fig:rl-xrl-dropped}
\end{figure}

We continue the performance evaluation of the proposed XRL approach by shedding light on the volume of dropped traffic, resulting in a violation of SLA requirements due to incorrect radio resource allocation policies. Fig.~\ref{fig:rl-xrl-dropped} shows the lopsided box plot of mMTC dropped traffic, and the results reveal that the maximum dropped traffic for the XRL scheme, excluding outliers, is \emph{5.2\%}, whereas it is \emph{7.9\%} for the RL method. Additionally, the XRL method experiences a positive skew, i.e., the mean value is greater than the median. In contrast, the RL method leads to a higher dropped traffic rate for the mMTC slice, where the median line of the RL box lies outside of the XRL box entirely, i.e., proves the difference between the two groups of distributed drop rate values and indicates the most perceived RL values are large.


\section{Conclusion}
In this paper, we have proposed a novel intrinsic interpretable multi-agent DRL as a practice to fulfill transparent SLA-aware 6G network slicing. We targeted the underlying resource allocation optimization task in a network slicing setup leveraging an explanation-guided DRL pipeline. The proposed scheme strives to learn optimal radio resource allocation under stringent SLA requirements such as latency. In particular, we proposed an XAI reward mechanism assisted by SHAP importance values over an extracted batch of state-action pairs and entropy mapper to encourage the agents to learn more relevant actions for specific network states. The numerical results validated the superiority of the XRL approach over the RL method concerning average return, XAI metric, latency, and dropped traffic.

\section*{Acknowledgment}
The research leading to these results has been partially supported by grant PID2021-126431OB-I00 funded by MCIN/AEI/ 10.13039/501100011033 and by "ERDF A way of making Europe", H2020 MonB5G Project (grant 871780), Program UNICO I+D under Grant TSI-063000- 2021-54/55, and Generalitat de Catalunya (grant 2021SGR-00770).

\end{document}